\documentclass[aps,preprint]{revtex4}%
\usepackage{amsfonts}
\usepackage{amsmath}
\usepackage{amssymb}
\usepackage{graphicx}%
\providecommand{\U}[1]{\protect\rule{.1in}{.1in}}
\newtheorem{theorem}{Theorem}
\newtheorem{acknowledgement}[theorem]{Acknowledgement}

\begin{document}
\title[Rotation curves]{Galaxy rotation curves. The theory.}
\author{Boris E. Meierovich}
\affiliation{P.L.Kapitza institute for physical problems 2 Kosygina str., Moscow 119334, Russia}
\begin{abstract}
The non-gauge vector field with as simple as possible Lagrangian
(\ref{Lagrangian}) turned out an adequate tool for macroscopic description of
the main properties of dark matter. The dependence of the velocity of a star
on the radius of the orbit $V\left(  r\right)  $ -- galaxy rotation curve --
is derived analytically from the first principles\ completely within the
Einstein's general relativity. The Milgrom's empirical modification of
Newtonian dynamics in nonrelativistic limit (MOND) gets justified and
specified in detail. In particular, the transition to a plateau is accompanied
by damping oscillations. In the scale of a galaxy, and in the scale of the
whole universe, the dark matter is described by a vector field with the same
energy-momentum tensor. It is the evidence of the common physical nature. Now,
when we have the general expression (\ref{Tik b=c=0}) for the energy-momentum
tensor of dark matter, it is possible to analyze its influence on the
structure and evolution of super heavy stars and black holes.

\end{abstract}
\maketitle


\section{Introduction}

The \textquotedblleft galaxy rotation curves\textquotedblright\ problem
appeared after J. H. Oort discovered the galactic halo, a group of stars
orbiting the Milky Way outside the main disk \cite{Oort 1924}. In 1933, F.
Zwicky \cite{Zwicky 1933} postulated "missing mass" to account for the orbital
velocities of galaxies in clusters. Persistent investigations by Vera Rubin
\cite{V Rubin} in seventies finally dispelled the skepticism about the
existence of dark matter on the periphery of the galaxies.

Among numerous attempts to solve the problem of galaxy rotation curves the
most discussed one is an empirical explanation named MOND (Modified Newtonian
Dynamics), proposed by Milgrom back in 1983 \cite{Milgrom}. For a relativistic
justification of MOND Bekenstein \cite{Bekenstein}, Sanders \cite{Sanders},
Brownstein and Moffat \cite{Brownstein},\cite{Moffat} introduce additional
scalar, vector, or tensor fields. Though these (and many others) relativistic
improvements of MOND are able to fit a large number of samples for about a
hundred galaxies, the concern still remains. So far we had neither
self-consistent description of the dark sector as a whole, nor direct
derivation of MOND from the first principles within the Einstein's general
relativity. The survey \cite{Fameay and McGaugh} by Benoit Famaey and Stacy
McGaugh reflects the current state of the research and contains the most
comprehensive list of references.

From my point of view the approach to the theory of dark sector based on the
utilization of vector fields in general relativity is very promising, and its
abilities are not yet exhausted. Vector fields with the simplest Lagrangian%
\begin{equation}
L=a\left(  \left(  \phi_{;K}^{K}\right)  ^{2}-m^{2}\phi^{K}\phi_{K}\right)
-V_{0} \label{Lagrangian}%
\end{equation}
allowed me to describe macroscopically the main features of evolution of the
Universe completely within the frames of the Einstein's theory of general
relativity \cite{Meierovich1}. The longitudinal non-gauge massive vector field
displays the repulsive elasticity. As a result the Big Bang turns into a
regular inflation-like state of maximum compression with the further
accelerated expansion at late times. The parametric freedom of the theory
allows to forget the fine tuning troubles. At the scales much larger than the
distances between the galaxies the Universe is homogeneous and isotropic. Its
temporal evolution depends on time only. Currently the characteristic rate of
its expansion is determined by the Hubble parameter which is of the order of
the inverse time from the big bang. In the much smaller galactic scales the
situation is just the opposite. The space structure is essentially
nonhomogeneous, while the influence of expansion is negligible.{}

In what follows I present the macroscopic theory of dark matter, including the
derivation of galaxy rotation curves, directly from the first principles
within the minimal Einstein's general relativity. In the galactic scale the
longitudinal non-gauge vector field with the same Lagrangian (\ref{Lagrangian}%
) not only fits the observed rotation curves, but also opens a promising
approach to understand the origin of the substance that we name dark matter.
In the nonrelativistic limit the expression (\ref{V (r)}) derived
analytically, justifies and specifies the empirical modification of Newtonian
dynamics by M. Milgrom \cite{Milgrom}.

\section{Vector field in general relativity}

In general relativity, the Lagrangian of a vector field $\phi_{I}$ consists of
the scalar bilinear combinations of its covariant derivatives and a scalar
potential $V(\phi^{K}\phi_{K})$. A bilinear combination of the covariant
derivatives is a 4-index tensor $S_{IKLM}=\phi_{I;K}\phi_{L;M}.$ The most
general form of the scalar $S$, formed via contractions of $S_{IKLM}$, is
$S=(ag^{IK}g^{LM}+bg^{IL}g^{KM}+cg^{IM}g^{KL})S_{IKLM},$ where $a,b,$ and $c$
are arbitrary constants. The general form of the Lagrangian of a vector field
$\phi_{I}$ is
\begin{equation}
L=a(\phi_{;M}^{M})^{2}+b\phi_{;M}^{L}\phi_{L}^{;M}+c\phi_{;M}^{L}\phi_{;L}%
^{M}-V(\phi_{M}\phi^{M}). \label{Lagrangian geeral}%
\end{equation}

The classification of vector fields $\phi_{I}$ is most convenient in terms of
the symmetric $G_{IK}=\frac{1}{2}(\phi_{I;K}+\phi_{K;I})$ and antisymmetric
$F_{IK}=\frac{1}{2}(\phi_{I;K}-\phi_{K;I})$ parts of the covariant
derivatives. The Lagrangian (\ref{Lagrangian geeral}) gets the form
\[
L=a(G_{M}^{M})^{2}+(b+c)G_{M}^{L}G_{L}^{M}+(b-c)F_{M}^{L}F_{\text{ }L}%
^{M}-V(\phi_{M}\phi^{M}).
\]
The bilinear combination of antisymmetric derivatives $F_{M}^{L}F_{\text{ }%
L}^{M}$ is the same as in electrodynamics. It becomes clear in the common
notations $A_{I}=\phi_{I}/2,$ $F_{IK}=A_{I;K}-A_{K;I}.$

The terms with symmetric covariant derivatives deserve special attention. In
applications of the vector fields to elementary particles in flat space-time
the divergence $\frac{\partial\phi^{K}}{\partial x^{K}}$ is artificially set
to zero \cite{Bogolubov Shirkov}:
\begin{equation}
\frac{\partial\phi^{K}}{\partial x^{K}}=0. \label{div fi =0}%
\end{equation}
This restriction allows to avoid the difficulty of negative contribution to
the energy. In the electromagnetic theory it is referred to as Lorentz gauge.
The negative energy problem in application to the galaxy rotation curves in
view of a precaution against instability of the vacuum was discussed by J.
Bekenstein \cite{Bekenstein}. However in general relativity (in curved
space-time) the energy is not a scalar, and its sign is not invariant against
the arbitrary coordinate transformations. From my point of view, considering
vector fields in general relativity, it is worth getting rid of the
restriction (\ref{div fi =0}), using instead a more weak condition of regularity.

The covariant field equations
\begin{equation}
a\phi_{;K;I}^{K}+b\phi_{I;K}^{;K}+c\phi_{;I;K}^{K}=-V^{\prime}\phi_{I}
\label{Covar field eqs}%
\end{equation}
and the energy-momentum tensor
\begin{equation}%
\begin{array}
[c]{l}%
T_{IK}=-g_{IK}L+2V^{\prime}\phi_{I}\phi_{K}+2ag_{IK}(\phi_{;M}^{M}\phi
^{L})_{;L}+2(b+c)[(G_{IK}\phi^{L})_{;L}-G_{K}^{L}F_{IL}-G_{I}^{L}F_{KL}]\\
+2(b-c)(2F_{\text{ \ }I}^{L}F_{LK}-F_{\text{ \ }K;L}^{L}\phi_{I}-F_{\text{
\ }I;L}^{L}\phi_{K})
\end{array}
\label{T_IK= general}%
\end{equation}
describe the behavior of a vector fields in the background of any arbitrary
given metric $g_{IK}$ \cite{Meierovich2}. Here $V^{\prime}\equiv\frac
{dV(\phi_{M}\phi^{M})}{d(\phi_{M}\phi^{M})}$.

If the back reaction of the field on the curvature of space-time is essential,
then the metric obeys the Einstein equations
\begin{equation}
R_{IK}-\frac{1}{2}g_{IK}R+\Lambda g_{IK}=\varkappa T_{IK}
\label{Einstein equations  General}%
\end{equation}
with (\ref{T_IK= general}) added to $T_{IK}.$ Here $\Lambda$ and $\varkappa$
are the cosmological and gravitational constants, respectively. With account
of back reaction the field equations (\ref{Covar field eqs}) are not
independent. They follow from the Einstein equations
(\ref{Einstein equations General}) with $T_{IK}$ (\ref{T_IK= general}) due to
the Bianchi identities. The field equations (\ref{Covar field eqs}) are linear
with respect to $\phi$ if the vector field is small, and the terms with the
second and higher derivatives of the potential $V\left(  \phi_{M}\phi
^{M}\right)  $ can be omitted.\bigskip

\section{Dark matter described by a vector field}

In curved space-time there is no invariance against the order of covariant
differentiation:%
\[
\phi_{;K;L}^{K}-\phi_{;L;K}^{K}=\phi^{M}R_{ML}.
\]
In general relativity there is no reason why the terms $\sim a$ in
(\ref{Lagrangian geeral}) and/or in (\ref{Covar field eqs}) should be "less
equal than others". In order to separate the dark matter from the ordinary one
it is reasonable to set $b=c=0.$ The case $a\neq0$ is supposed to describe the
dark matter only. The opposite case $a=0,$ and $b\neq0,c\neq0$ corresponds to
either electromagnetic field $\left(  c=-b\right)  ,$ or to vector particles
$\left(  b\neq0,c=0\right)  .$ This way the dark matter and the ordinary
matter are separated from one another, so that the ordinary matter is not
taken into account twice. The dark matter is described by the Lagrangian
\begin{equation}
L_{\text{dm}}=a(\phi_{;M}^{M})^{2}-V(\phi_{M}\phi^{M}).
\label{dark matter Lagrangian}%
\end{equation}

Thereafter the field equation (\ref{Covar field eqs}) and the energy-momentum
tensor of the vector field (\ref{T_IK= general}) reduce to%
\begin{align}
a\frac{\partial\phi_{;M}^{M}}{\partial x^{I}}  &  =-V^{\prime}\phi
_{I},\label{Field eqs b=c=0}\\
T_{\text{dm }IK}  &  =g_{IK}\left[  (\phi_{;M}^{M})^{2}/a+V\right]
+2V^{\prime}\left(  \phi_{I}\phi_{K}-g_{IK}\phi^{M}\phi_{M}\right)  .
\label{Tik b=c=0}%
\end{align}
Though the dark matter displays itself by curving the space-time, its physical
nature remains unclear so far. We don't know the dependence $V(\phi_{M}%
\phi^{M}).$ If the vector $\phi_{I}$ remains small enough to neglect the
second and higher derivatives of $V(\phi_{M}\phi^{M}),$ then the parameter
\[
m^{2}=\left\vert \frac{V^{\prime}\left(  0\right)  }{a}\right\vert
\]
characterizes the field. As usual it\ is designated as the square of mass. In
accordance with (\ref{Field eqs b=c=0}) the dimension of $m$ is $cm^{-1}$. The
covariant divergence $\phi_{;M}^{M}$\ is a scalar, and in accordance with the
equation (\ref{Field eqs b=c=0}) the massive $\left(  m\neq0\right)  $ field
has a potential: it is a gradient of a scalar.

So far there is no evidence of any direct interaction between dark and
ordinary matter other than via gravitation. The gravitational interaction is
described by Einstein equations (\ref{Einstein equations General}) with
\begin{equation}
T_{IK}=T_{\text{dm }IK}+T_{\text{om }IK}, \label{T_IK=dm+om}%
\end{equation}
where
\begin{equation}
T_{\text{om }IK}=\left(  \varepsilon+p\right)  u_{I}u_{K}-pg_{IK}
\label{Tom_IK}%
\end{equation}
is the well known energy-momentum tensor of macroscopic objects. The energy
$\varepsilon,$ pressure $p,$ and temperature $T$ of the ordinary matter obey
the equation of state. If $T\ll\varepsilon$ the Einstein equations
(\ref{Einstein equations General}) with $T_{IK}$ (\ref{T_IK=dm+om}) together
with the equation of state with $T=0$ form a complete set. The field equation
(\ref{Field eqs b=c=0}) is not independent. It is a consequence of the
Einstein equations due to Bianci identities.

\section{Galaxy rotation curves}

Applying general relativity to the galaxy rotation problem it is reasonable to
consider a static centrally symmetric metric%

\begin{equation}
ds^{2}=g_{IK}dx^{I}dx^{K}=e^{\nu\left(  r\right)  }\left(  dx^{0}\right)
^{2}-e^{\lambda\left(  r\right)  }dr^{2}-r^{2}d\Omega^{2}
\label{Centr symm metric}%
\end{equation}
with two functions $\nu\left(  r\right)  $ and $\lambda\left(  r\right)  $
depending on only one coordinate - circular radius $r$. Real distribution of
the stars and planets in a galaxy is neither static, nor centrally symmetric.
However this simplification facilitates analyzing the problem and allows to
display the main results analytically. If a galaxy is concentrated around a
supermassive black hole, the deviation from the central symmetry caused by the
peripheral stars is small.

In the background of the centrally symmetric metric (\ref{Centr symm metric})
the vector $\phi^{I}$ is longitudinal. In accordance with the field equation
(\ref{Field eqs b=c=0}) its only non-zero component $\phi^{r}$ depends on $r.$
In view of
\[
g=\det g_{IK}=-e^{\lambda+\nu}r^{4}\sin^{2}\theta,\qquad\frac{1}{\sqrt{-g}%
}\frac{\partial\sqrt{-g}}{\partial r}=\frac{2}{r}+\frac{\lambda^{\prime}%
+\nu^{\prime}}{2}.
\]
the covariant divergence%
\begin{equation}
\phi_{;M}^{M}=\frac{1}{\sqrt{-g}}\frac{\partial\left(  \sqrt{-g}\phi
^{M}\right)  }{\partial x^{M}}=\frac{\partial\phi^{r}}{\partial r}+\left(
\frac{2}{r}+\frac{\lambda^{\prime}+\nu^{\prime}}{2}\right)  \phi
^{r}.\label{div(fi)}%
\end{equation}

In the \textquotedblleft dust matter\textquotedblright\ approximation $p=0,$
and the only nonzero component of the energy-momentum tensor (\ref{Tom_IK}) is
$T_{\text{om }00}=\varepsilon g_{00}$. Whatever the distribution of the
ordinary matter $\varepsilon\left(  r\right)  $ is, the covariant divergence
$T_{\text{om }I;K}^{K}$ is automatically zero. In the dust
matter\ approximation the curving of space-time by ordinary matter is taken
into account, but the back reaction of the gravitational field on the
distribution of matter is ignored. If $p=0$ the energy $\varepsilon\left(
r\right)  $ is considered as a given function.

In the power series
\[
V(\phi_{M}\phi^{M})=V_{0}+V^{\prime}\phi_{M}\phi^{M}+O\left(  \left(  \phi
_{M}\phi^{M}\right)  ^{2}\right)
\]
$V_{0}=V\left(  0\right)  $ together with the cosmological constant $\Lambda$
determines the expansion of the Universe. In the scale of galaxies the role of
expansion of the Universe as a whole is negligible, and one can set
$\widetilde{\Lambda}=\Lambda-\varkappa V_{0}=0$ in the Einstein equations.
Omitting the second and higher derivatives of the potential $V(\phi_{M}%
\phi^{M}),$ we have the Einstein equations as follows (see
\cite{Landau-Lifshits}, page 382 for the derivation of the left-hand sides):%
\begin{align}
-e^{-\lambda}\left(  \frac{1}{r^{2}}-\frac{\lambda^{\prime}}{r}\right)
+\frac{1}{r^{2}}  &  =\varkappa T_{0}^{0}=\varkappa\left[  (\phi_{;M}^{M}%
)^{2}/a+V^{\prime}e^{\lambda}\left(  \phi^{r}\right)  ^{2}+\varepsilon\right]
\label{0-0 Eq}\\
-e^{-\lambda}\left(  \frac{\nu^{\prime}}{r}+\frac{1}{r^{2}}\right)  +\frac
{1}{r^{2}}  &  =\varkappa T_{r}^{r}=\varkappa\left[  (\phi_{;M}^{M}%
)^{2}/a-V^{\prime}e^{\lambda}\left(  \phi^{r}\right)  ^{2}-p\right]
\label{r-r Eq}\\
-\frac{1}{2}e^{-\lambda}\left(  \nu^{\prime\prime}+\frac{\nu^{\prime2}}%
{2}+\frac{\nu^{\prime}-\lambda^{\prime}}{r}-\frac{\nu^{\prime}\lambda^{\prime
}}{2}\right)   &  =\varkappa\left[  (\phi_{;M}^{M})^{2}/a+V^{\prime}%
e^{\lambda}\left(  \phi^{r}\right)  ^{2}-p\right]  ,\qquad I,K\neq0,r.
\label{fi-fi and teta-teta Eqs}%
\end{align}
Here prime $^{\prime}$ stands for $\frac{d}{dr},$ except $V^{\prime}%
=\frac{\partial V\left(  \phi_{M}\phi^{M}\right)  }{\partial\left(  \phi
_{M}\phi^{M}\right)  }.$ Among the four equations (\ref{Field eqs b=c=0}),
(\ref{0-0 Eq}-\ref{fi-fi and teta-teta Eqs}) for the unknowns $\phi
^{r},\lambda,$ and $\nu$ any three are independent.

Extracting (\ref{r-r Eq}) from (\ref{0-0 Eq}) we get a relation%
\begin{equation}
\nu^{\prime}+\lambda^{\prime}=\varkappa re^{\lambda}\left[  2e^{\lambda
}\left(  \phi^{r}\right)  ^{2}V^{\prime}+\varepsilon+p\right]  .
\label{Equation for nu'+lambda'}%
\end{equation}
With account of (\ref{div(fi)}) and (\ref{Equation for nu'+lambda'}) the
vector field equation (\ref{Field eqs b=c=0}) takes the form%
\begin{equation}
\left[  \left(  \phi^{r}\right)  ^{\prime}+\left(  \frac{2}{r}+\varkappa
re^{2\lambda}\left(  \phi^{r}\right)  ^{2}V^{\prime}+\frac{1}{2}\varkappa
re^{\lambda}\left(  \varepsilon+p\right)  \right)  \phi^{r}\right]  ^{\prime
}=-m^{2}e^{\lambda}\phi^{r} \label{Field eq spherical symm}%
\end{equation}
where $m^{2}=-\frac{V^{\prime}\left(  0\right)  }{a}.$ The sign in the r.h.s.
of (\ref{Field eq spherical symm}) corresponds to the case $V^{\prime}\left(
0\right)  >0,$ $a<0.$ Negative $a$ is taken in accordance with the
requirements of regularity in application of the same Lagrangian
(\ref{dark matter Lagrangian}) to the analysis of the role of dark matter in
the evolution of the Universe \cite{Meierovich1}. It is convenient to set
$a=-1$ in what follows. Hence $V^{\prime}\left(  0\right)  =m^{2}.$ The
equations (\ref{Equation for nu'+lambda'}) and (\ref{Field eq spherical symm})
are derived with no assumptions concerning the strength of the gravitational field.

Excluding $\lambda^{\prime}$ from equations (\ref{0-0 Eq}) and
(\ref{Equation for nu'+lambda'}), we get the following expression for
$\nu^{\prime}:$
\[
\nu^{\prime}=\varkappa re^{\lambda}\left[  m^{2}e^{\lambda}\left(  \phi
^{r}\right)  ^{2}+(\phi_{;M}^{M})^{2}+p\right]  +\frac{e^{\lambda}-1}{r}%
\]

In case of the dust matter approximation $\left(  p=0\right)  :$
\begin{equation}
\nu^{\prime}=\varkappa re^{\lambda}\left[  m^{2}e^{\lambda}\left(  \phi
^{r}\right)  ^{2}+(\phi_{;M}^{M})^{2}\right]  +\frac{e^{\lambda}-1}{r}
\label{nu'=-lambda' +...}%
\end{equation}
\ In a static centrally symmetric gravitational field$\ \nu^{\prime}$
determines the centripetal acceleration of a particle (\cite{Landau-Lifshits},
page 323). Without dark matter $\phi^{r}=0$ (\ref{nu'=-lambda' +...}) gives
the Newton's attractive potential far from the center:
\[
\varphi_{\text{N}}\left(  r\right)  =\frac{1}{2}c^{2}\nu\left(  r\right)
\sim-r^{-1},\qquad r\rightarrow\infty.
\]
The first term in the r.h.s. of (\ref{nu'=-lambda' +...}) appears due to the
dark matter. Both terms have the same sign, and the presence of dark matter
increases the attraction to the center.

The curvature of space-time caused by a galaxy is small. In the linear
approximation the influence of dark and ordinary matter can be separated from
one another. For $\lambda\ll1$ (\ref{nu'=-lambda' +...}) reduces to
\begin{equation}
\nu^{\prime}=\varkappa r\left[  m^{2}\left(  \phi^{r}\right)  ^{2}+(\phi
_{;M}^{M})^{2}\right]  +\frac{\lambda}{r}, \label{nu'=...+lambda/r}%
\end{equation}
where the first term does not contain $\varepsilon.$ However, the contribution
of dark matter comes from both additives.\ The vector field equation
(\ref{Field eq spherical symm}) and the Einstein equation (\ref{0-0 Eq}) at
$\lambda\ll1$ are simplified:%
\begin{equation}
\left(  \phi^{r}\right)  ^{\prime\prime}+\left(  \left[  \frac{2}{r}+\varkappa
m^{2}r\left(  \phi^{r}\right)  ^{2}+\frac{1}{2}\varkappa r\left(
\varepsilon+p\right)  \right]  \phi^{r}\right)  ^{\prime}=-m^{2}\phi^{r}
\label{vec field eq lambda very small}%
\end{equation}%
\begin{equation}
\lambda^{\prime}+\frac{\lambda}{r}=\varkappa r\left[  -(\phi_{;M}^{M}%
)^{2}+m^{2}\left(  \phi^{r}\right)  ^{2}+\varepsilon\right]  .
\label{Eq for lambda nonrelativistic}%
\end{equation}
The boundary conditions for these equations,
\begin{equation}
\phi^{r}=\frac{1}{3}\phi_{0}^{\prime}r,\qquad\lambda=\frac{1}{3}%
\varkappa\left(  \varepsilon_{0}-\phi_{0}^{\prime2}\right)  r^{2},\qquad
r\rightarrow0, \label{Boundary conditions}%
\end{equation}
are determined by the requirement of regularity in the center. Here
$\varepsilon_{0}=\varepsilon\left(  0\right)  .$

The term $\frac{1}{2}\varkappa r\left(  \varepsilon+p\right)  $ in
(\ref{vec field eq lambda very small})\ reflects the interaction of dark and
ordinary matter via gravitation. If the curvature of space-time caused by the
ordinary matter is small, this term is negligible compared to $2/r.$ The
nonlinear term $\varkappa m^{2}r\left(  \phi^{r}\right)  ^{2}$ is small
compared to $2/r$ at $r\rightarrow0,$ but at $r\rightarrow\infty,$ despite of
being small, it decreases only a little bit quicker than $2/r$\footnote[1]%
{This nonlinear term at $r\rightarrow\infty$ decreases as $\left(  r\ln
r\right)  ^{-1}:$
\[
\varkappa m^{2}r\left(  \phi^{r}\right)  ^{2}=\frac{2\sin^{2}mr}{3r\ln\frac
{r}{r^{\ast}}}\approx\frac{1}{3r\ln\frac{r}{r^{\ast}}},\qquad r^{\ast}%
\sim\frac{1}{m}.
\]
}$.$ Neglecting both nonlinear terms in square brackets, the field equation
(\ref{vec field eq lambda very small}) reduces to%
\[
\left(  \left(  \phi^{r}\right)  ^{\prime}+\frac{2}{r}\phi^{r}\right)
^{\prime}=-m^{2}\phi^{r}.
\]
Its regular solution is
\begin{equation}
\phi^{r}=\frac{\phi_{0}^{\prime}}{m^{3}r^{2}}\left(  \sin mr-mr\cos mr\right)
,\qquad\phi_{;M}^{M}=\phi_{0}^{\prime}\frac{\sin mr}{mr},
\label{analytic fi and fi;m}%
\end{equation}
where $\phi_{0}^{\prime}=\phi_{;M}^{M}\left(  0\right)  .$ Substitution of
(\ref{analytic fi and fi;m}) into (\ref{nu'=...+lambda/r}) results in%
\[
\nu^{\prime}\left(  r\right)  =\frac{\varkappa(\phi_{0}^{\prime})^{2}}{m^{2}%
r}f\left(  mr\right)  +\frac{\lambda}{r},\qquad\lambda\ll1.
\]
Function $f\left(  x\right)  ,$
\begin{equation}
f\left(  x\right)  =\left(  1-\frac{\sin2x}{x}+\frac{\sin^{2}x}{x^{2}}\right)
=\left\{
\begin{array}
[c]{c}%
\allowbreak x^{2}-\frac{2}{9}x^{4}+...,\qquad x\rightarrow0\\
1,\qquad x\rightarrow\infty
\end{array}
\right.  , \label{f(x) analytic}%
\end{equation}
is presented in Figure 1 (blue curve).%
\[%
{\parbox[b]{2.3744in}{\begin{center}
\includegraphics[
height=1.666in,
width=2.3744in
]%
{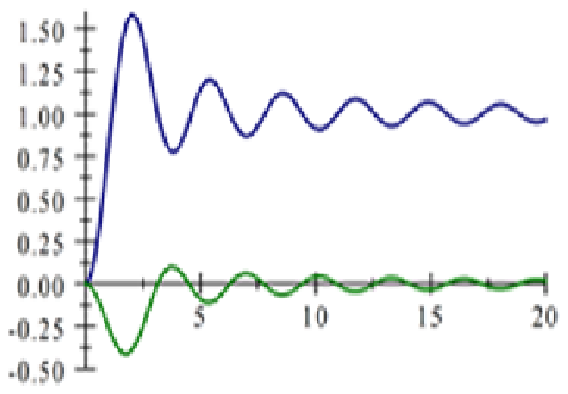}%
\\
Figure 1. Function $f\left(  x\right)  $ (\ref{f(x) analytic}) - blue curve,
and $\Psi\left(  x\right)  $ (\ref{Psi(x)}) - green curve.
\end{center}}}
\qquad%
{\parbox[b]{1.9898in}{\begin{center}
\includegraphics[
height=1.5848in,
width=1.9898in
]%
{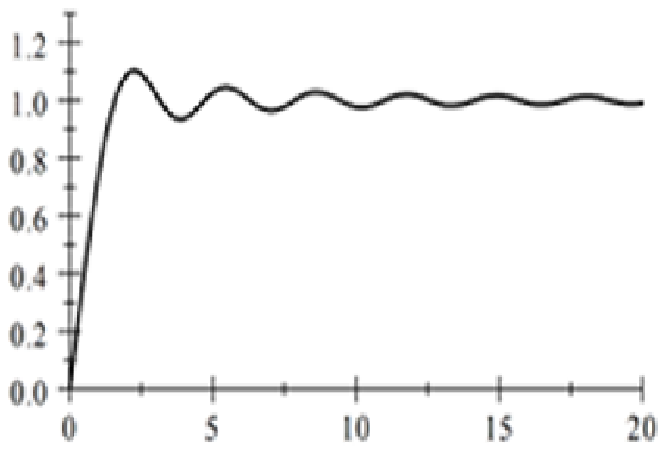}%
\\
Figure 2. Function $\sqrt{f\left(  x\right)  +\Psi\left(  x\right)  }$ in
\ (\ref{V(r) at eps=0})\ found analytically coinsides with the found
numerically.
\end{center}}}
\qquad
\]

The balance of the centripetal $\frac{c^{2}\nu^{\prime}}{2}$ and centrifugal
$\frac{V^{2}}{r}$ accelerations determines the velocity $V$ of a rotating
object as a function of the radius $r$ of its orbit:%
\begin{align}
V\left(  r\right)   &  =\sqrt{V_{\text{pl}}^{2}f\left(  mr\right)  \dot
{+}\frac{c^{2}}{2}\lambda\left(  r\right)  },\label{V (r)}\\
V_{\text{pl}} &  =\sqrt{\frac{\varkappa}{2}}\frac{c\phi_{0}^{\prime}}%
{m}\label{Velocity on plateau}%
\end{align}
Far from the center $\lambda\left(  r\right)  $ decreases as $1/r,$ while
$f\left(  mr\right)  \rightarrow1.$ The dependence $V\left(  r\right)  $
(\ref{V (r)}) turns at $r\gtrsim m^{-1}$ from a linear to a plateau with
damping oscillations. The plateau appears entirely due to the vector field. At
the same time the vector field contributes to $\lambda\left(  r\right)  $ as
well. Regular at $r\rightarrow0$ solution of the equation
(\ref{Eq for lambda nonrelativistic}) is%

\begin{equation}
\lambda\left(  r\right)  =2\left(  \frac{V_{\text{pl}}}{c}\right)  ^{2}%
\Psi\left(  mr\right)  +\frac{\varkappa}{r}\int_{0}^{r}\varepsilon\left(
r\right)  r^{2}dr,\qquad\lambda\ll1. \label{small lambda(r)}%
\end{equation}
The last term in (\ref{small lambda(r)}) gives the Newton's potential. The
function
\begin{equation}
\Psi\left(  x\right)  =\frac{1}{x}\int_{0}^{x}\left(  \frac{\sin^{2}y}{y^{2}%
}-\frac{\sin2y}{y}+\cos2y\right)  dy \label{Psi(x)}%
\end{equation}
is shown in Figure 1 (green curve). The radial dependence $V\left(  r\right)
/V_{\text{pl}}$ at $\varepsilon\rightarrow0,$
\begin{equation}
V\left(  r\right)  /V_{\text{pl}}=\sqrt{f\left(  mr\right)  +\Psi\left(
mr\right)  }, \label{V(r) at eps=0}%
\end{equation}
is shown in Figure 2. In the limit $\lambda\ll1$ the transition to a plateau
due to the dark matter only is a universal function (\ref{V(r) at eps=0}).

The plateau value $V_{\text{pl}}$ (\ref{Velocity on plateau}) is connected
with the parameter $\phi_{0}^{\prime}/m,$ and the period of oscillations is
$\frac{2\pi}{m}.$ The form of a plateau\ allows to restore the value of the
parameter $\phi_{0}^{\prime}=\phi_{;M}^{M}\left(  0\right)  $ at
$r\rightarrow0$ in the boundary conditions (\ref{Boundary conditions})$.$ As
far as there is no evidence of any direct interaction of dark and ordinary
matter, the origin of specific values $\phi_{0}^{\prime}$ and $m$ of a
particular galaxy depends on what happens in the center. As long as the
gravitation is weak, in the linear approximation $\phi_{0}^{\prime},$
$\varepsilon_{0},$ and $m$ are free parameters. The values $V_{\text{pl}}$ and
$m$ can differ from one galaxy to another. It looks like for each galaxy the
values of $V_{\text{pl}}$ and $m$ are driven by some heavy object (may be a
black hole, may be a neutron star) located in the center (by the way,
supporting the central symmetry of the gravitational field).

Interaction with dark matter via gravitation should affect the equilibrium
structure of heavy stars and can shift the collapse boundary. The Einstein
equations are not linear. If the gravitation is not weak, there are
restrictions on the parameters $\phi_{0}^{\prime}$ and $\varepsilon_{0}$ in
(\ref{Boundary conditions}). If $p\neq0$ the radial distribution of the
ordinary matter and gravitational field are interdependent. In the
approximation of cold degenerate relativistic gas it is more convenient to use
the chemical potential $\mu_{0}$ in the boundary conditions instead of
$\varepsilon_{0}.$ It is worth reconsidering the equilibrium \cite{Oppenheimer
Volkov} and collapse \cite{Oppenheimer Snyder} of supermassive bodies taking
the dark matter into account. However, it is a different story.

Dark matter, described by a vector field with the Lagrangian (\ref{Lagrangian}%
), actually justifies the empirical Milgrom's hypothesis of MOND - the
modified Newton's dynamics \cite{Milgrom}. Naturally, basing only on the
intuitive arguments, it was scarcely possible to guess that the transition to
a plateau is accompanied by damping oscillations.

\section{Fitting}

The field itself is zero in the center, $\phi^{r}\left(  0\right)  =0$, and
the contributions of the dark matter and of the ordinary one are introduced to
the boundary conditions (\ref{Boundary conditions}) by the values $\left(
\phi_{0}^{\prime}\right)  ^{2}$ {}{}and $\varepsilon_{0},$ respectively.\ 

When there is a plateau, the speed of rotation on the plateau $V_{\text{pl}}$
(\ref{Velocity on plateau}), which is determined from the Doppler shift of
spectral lines, provides us with information about the input parameter
$\phi_{0}^{\prime}=\phi_{;M}^{M}\left(  0\right)  $ in the boundary
conditions. The parameter $m$ is determined by scaling the radial coordinate
so that the period of oscillations of $f\left(  x\right)  $
(\ref{f(x) analytic}) fits the observations. While the distribution of dark
matter is characterized unambiguously by the two parameters $\phi_{0}^{\prime
}$\ and $m,$ the situation with the density of the ordinary matter
$\varepsilon\left(  r\right)  $ in galaxies is not that clear. Radiation
coming from the galaxies does not carry information about cooled non-emitting
stars and planets. Just the opposite: the strict fitting could provide us with
the distribution of the ordinary matter in galaxies.

In the dust matter approximation and weak gravitational field $\varepsilon
\left(  r\right)  $ is an arbitrary given function. To demonstrate the
relative role of dark and ordinary matter I use the Gauss distribution for the
density of dust matter%
\begin{equation}
\varepsilon\left(  r\right)  =\varepsilon_{0}\exp\left(  -r^{2}/r_{0}%
^{2}\right)  . \label{eps(r) Gaus}%
\end{equation}
Qualitatively a particular form of a monotonically decreasing function
$\varepsilon\left(  r\right)  $ is not essential. (Possible existence of a
hard core in the center is a special case.) $\varepsilon_{0}$ is the maximum
density in the center, and $r_{0}$ is the mean radius of a galaxy. Total mass
of a galaxy $M\sim\varepsilon_{0}r_{0}^{3}.$ Though the dark and ordinary
matter are inputted into the boundary conditions (\ref{Boundary conditions})
via $\phi_{0}^{\prime}$ and $\varepsilon_{0}=\varepsilon\left(  0\right)  ,$
it looks more clearly to demonstrate their relative role using $\varepsilon
_{0}r_{0}^{3}$ (proportional to the total rest energy of a galaxy) instead of
$\varepsilon_{0}.$

In Figures 3 and 4 a blue dashed curve is the rotation curve
(\ref{V(r) at eps=0}) without ordinary matter. It is the same curve as in
Figure 2. In each case the radial scales are specifically chosen to clarify
the difference better. Red lines in figures 3 a,b,c are rotation curves with
$m^{2}r_{0}^{2}=1,10,0.1,$ respectively (the three cases where the radius
$r_{0}$ of a galaxy is equal, $\sqrt{10}$\ times larger, and $\sqrt{10}%
$\ times smaller then the period $\sim m^{-1}$ of oscillations). The ratio
$\frac{\varepsilon_{0}r_{0}^{3}}{\left(  \phi_{0}^{\prime}\right)  ^{2}}=1.$
The smaller is $\frac{\varepsilon_{0}r_{0}^{3}}{\left(  \phi_{0}^{\prime
}\right)  ^{2}},$ the less is the difference between red and blue curves.%
\[%
{\parbox[b]{1.9668in}{\begin{center}
\includegraphics[
height=1.3184in,
width=1.9668in
]%
{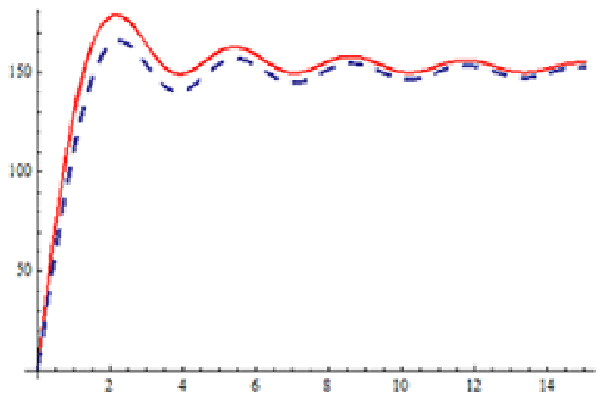}%
\\
Figure 3a. Red curve - (\ref{V (r)}) with $\frac{\varepsilon_0r_0^3}{\left(
\phi_0^\prime\right)  ^2}=1,$ \ \ $m^2r_0^2=1.$ Blue dashed curve -
(\ref{V(r) at eps=0}).
\end{center}}}
\qquad%
{\parbox[b]{1.9751in}{\begin{center}
\includegraphics[
height=1.3234in,
width=1.9751in
]%
{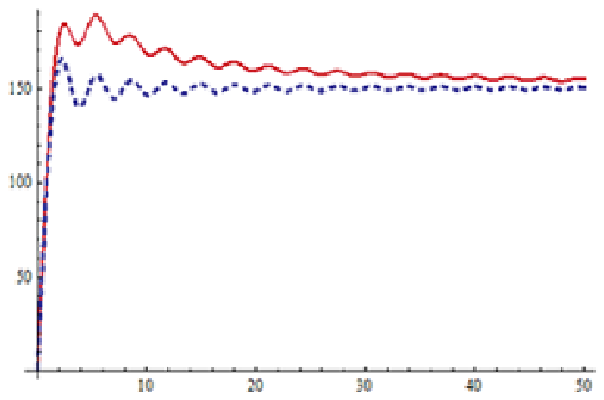}%
\\
Figure 3b. Red curve - (\ref{V (r)}) with $\frac{\varepsilon_0r_0^3}{\left(
\phi_0^\prime\right)  ^2}=1,$ \ \ $m^2r_0^2=10.$ Blue dashed curve -
(\ref{V(r) at eps=0}).
\end{center}}}
\qquad\
{\parbox[b]{1.9842in}{\begin{center}
\includegraphics[
height=1.32in,
width=1.9842in
]%
{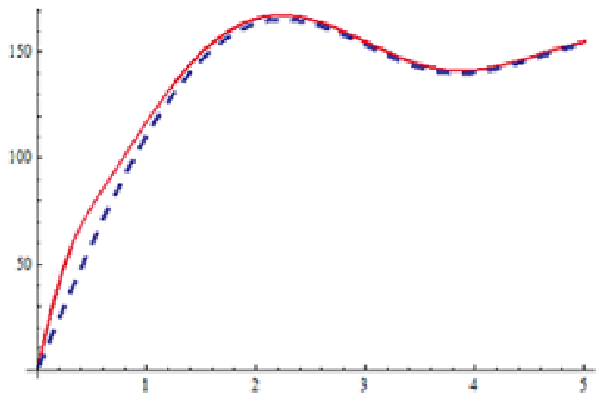}%
\\
Figure 3c. Red curve - (\ref{V (r)}) with $\frac{\varepsilon_0r_0^3}{\left(
\phi_0^\prime\right)  ^2}=1,$ \ \ $m^2r_0^2=0.1.$ Blue - (\ref{V(r) at eps=0}%
).
\end{center}}}
\]

Red curves in figures 4 a,b,c are rotation curves for a fixed $\frac
{\varepsilon_{0}r_{0}^{3}}{\left(  \phi_{0}^{\prime}\right)  ^{2}}=10,$ and
$m^{2}r_{0}^{2}=1,10,$ and $0.1,$ respectively. As $m^{2}r_{0}^{2}$ grows, the
oscillations are smoothed out, and when it decreases the difference between
the curves moves to the center.
\[%
{\parbox[b]{1.9676in}{\begin{center}
\includegraphics[
height=1.3164in,
width=1.9676in
]%
{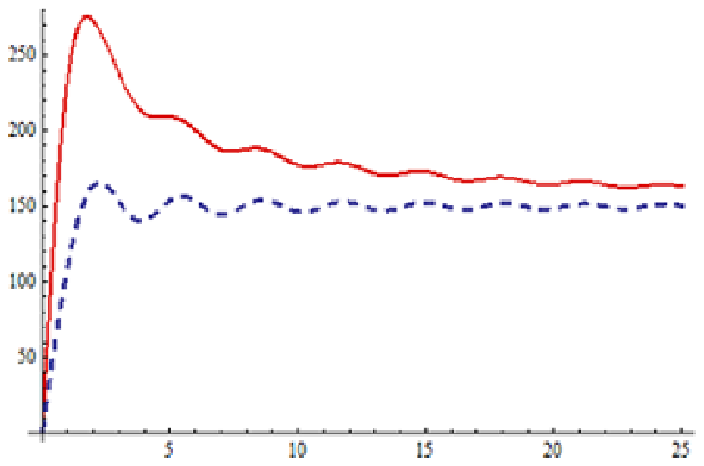}%
\\
Figure 4a.\ Red curve - (\ref{V (r)}) with $\frac{\varepsilon_0r_0^3}{\left(
\phi_0^\prime\right)  ^2}=10,$ \ \ $m^2r_0^2=1.$ Blue dashed curve -
(\ref{V(r) at eps=0}).
\end{center}}}
\qquad%
{\parbox[b]{1.9593in}{\begin{center}
\includegraphics[
height=1.3117in,
width=1.9593in
]%
{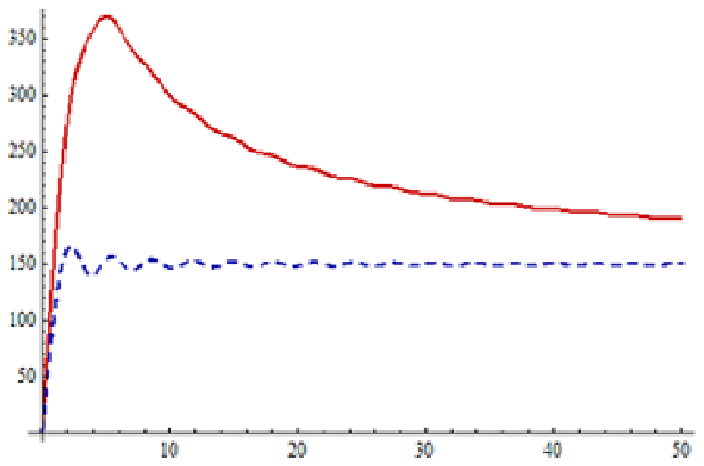}%
\\
Figure 4b. Red curve - (\ref{V (r)}) with $\frac{\varepsilon_0r_0^3}{\left(
\phi_0^\prime\right)  ^2}=10,$ \ \ $m^2r_0^2=10.$ Blue dashed curve -
(\ref{V(r) at eps=0}).
\end{center}}}
\text{\qquad}%
{\parbox[b]{1.9909in}{\begin{center}
\includegraphics[
height=1.3234in,
width=1.9909in
]%
{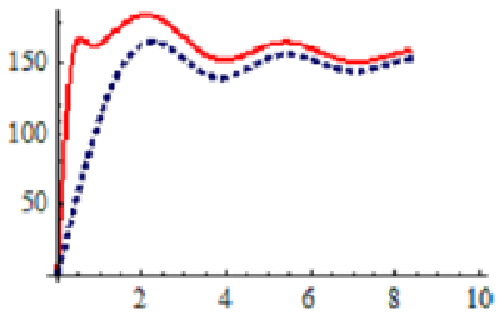}%
\\
Figure 4c. Red curve - (\ref{V (r)}) with $\frac{\varepsilon_0r_0^3}{\left(
\phi_0^\prime\right)  ^2}=10,$ \ \ $m^2r_0^2=0.1.$ Blue dashed curve -
(\ref{V(r) at eps=0}).
\end{center}}}
\]

One can find over a hundred graphs of galaxy rotation curves in the
literature, including those displaying the transition to a plateau. One of the
often referred to, marked UMa: NGC 3726, is shown in Figure 5$a$ and 5$b$.
Both graphs are taken from different places within the same list
\cite{Brownstein}.
\[%
{\parbox[b]{1.8489in}{\begin{center}
\includegraphics[
height=1.5774in,
width=1.8489in
]%
{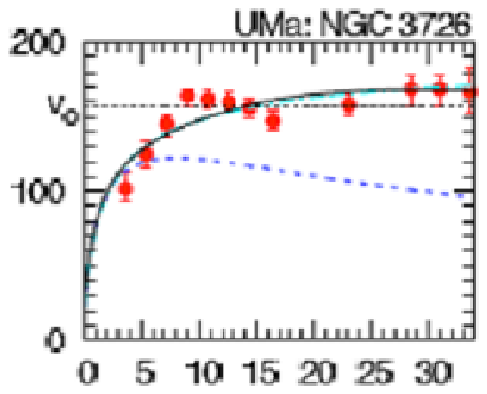}%
\\
Figure 5$a$. Fitting by MSTG, practically coinsiding with MOND.
\end{center}}}
\qquad%
{\parbox[b]{1.9909in}{\begin{center}
\includegraphics[
height=1.5733in,
width=1.9909in
]%
{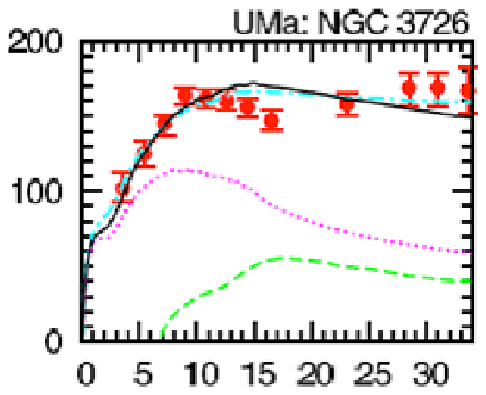}%
\\
Figure 5$b.$Another fitting by MSTG, slightly different from MOND.\ \ \
\end{center}}}
\qquad%
{\parbox[b]{2.1328in}{\begin{center}
\includegraphics[
height=1.5807in,
width=2.1328in
]%
{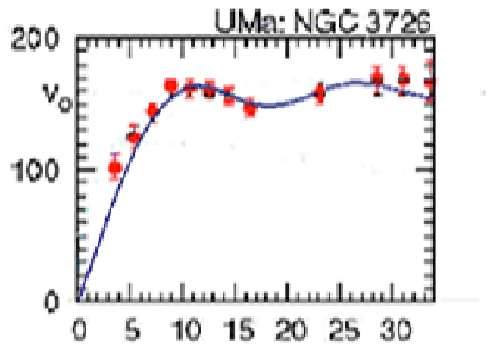}%
\\
Figure 5$c$. Points are observations, fitted by (\ref{V (r)}) together with
(\ref{small lambda(r)}) and (\ref{eps(r) Gaus}).
\end{center}}}
\]
The red points with error bars are the observations. The solid lines are the
rotation curves determined from so called MSTG\footnote[2]{MSTG is an attempt
to \textquotedblleft generalize\textquotedblright\ the general relativity on
the basis of a pseudo-Riemannian metric tensor and a skew symmetric rank-3
tensor field in a hope to explain the flat rotation curves of
galaxies\ \cite{Moffat}.} (metric-skew-tensor-gravity (\cite{Brownstein}%
,\cite{Moffat})). The solid line in Figure 5$a$ practically coincides with
MOND (modified Newton's dynamics). In Figure 5$b$ it is slightly different
from MOND. Other dashed and dotted lines correspond to the ordinary Newton's dynamics.

The blue curve in Figure 5$c$ shows how (\ref{V (r)}) together with
(\ref{small lambda(r)}) and (\ref{eps(r) Gaus}) fits the observations.\ The
input parameters are $V_{\text{pl}}=158$ $\frac{\text{Km}}{\sec},$
$\frac{\varkappa\varepsilon_{0}}{m^{2}}$ $=0.00000005$, $mr_{0}$ $=3.78$.

Figure 6 is another example of comparison of fitting by MSTG - MOND $\left(
a\right)  $ and in accordance with (\ref{V (r)})-(\ref{small lambda(r)}%
)-(\ref{eps(r) Gaus}) $\left(  b\right)  $. The input parameters are
$V_{\text{pl}}=130$ $\frac{\text{Km}}{\sec},$ $\frac{\varkappa\varepsilon_{0}%
}{m^{2}}$ $=0.00000005$, $mr_{0}$ $=2.24$.%

\[%
{\parbox[b]{2.2773in}{\begin{center}
\includegraphics[
height=1.8182in,
width=2.2773in
]%
{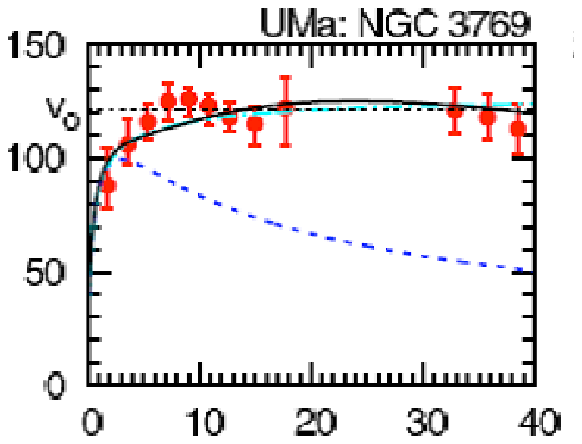}%
\\
Figure 6$a.$ Solid line is fitting via MSTG and MOND. Dashed -- the Newtons's
dynamics.
\end{center}}}
\qquad%
{\parbox[b]{2.5413in}{\begin{center}
\includegraphics[
height=1.8107in,
width=2.5413in
]%
{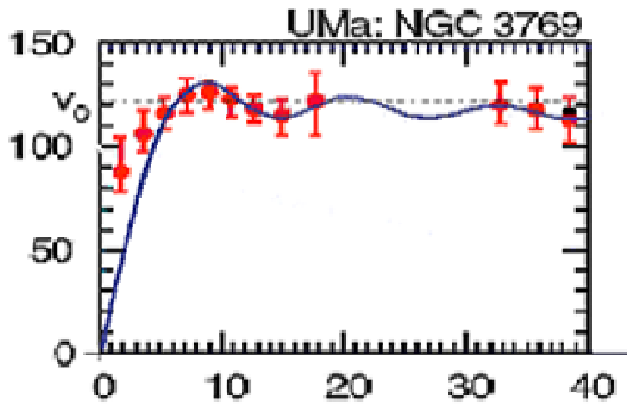}%
\\
Figure 6$b.$ Points are observations, Solid curve is fitting by (\ref{V (r)})
together with (\ref{small lambda(r)}) and (\ref{eps(r) Gaus}).
\end{center}}}
\]
\qquad

Frankly speaking, it is a surprise for me. I did not expect such a
coincidence. Deviations at small radii can be related to the presence of an
additional strongly gravitating compact object located at the center. As shown
in Figures 3$c$ and 4$c$ in case $r_{0}\ll m^{-1}$ the initial growing part of
the curve $V\left(  r\right)  $ shifts to the center. At the same time, if a
heavy object in the center really exists, it supports the central symmetry,
and the gravitational field becomes only slightly distorted by other stars and
planets of the galaxy.

\section{Summary}

The non-gauge vector field with as simple as possible Lagrangian
(\ref{Lagrangian}) provides the macroscopic description of all major observed
properties of the dark sector within the Einstein's theory of general relativity.

In the galaxy scale the field with the energy-momentum tensor (\ref{Tik b=c=0}%
) allows to describe analytically the galaxy rotation curves in detail. The
formulae (\ref{V (r)}$-$\ref{Psi(x)}) are derived completely within the
Einstein's theory. Thus, there is no need in any modifications of the general
relativity to explain the observable plateau in rotation curves.

As I have shown previously \cite{Meierovich1}, the vector fields with the same
Lagrangian (\ref{Lagrangian}) are adequate tools for macroscopic description
of the main features of evolution of the universe. In the scale of the whole
universe the zero-mass field corresponds to the dark energy, and the massive
one - to dark matter. Price issue is the rejection of the prejudice (widely
spread, unfortunately) that the energy should not be negative. Instead I
utilized a weaker condition of regularity. In general relativity the energy is
not a scalar, and its sign is not invariant against the arbitrary coordinate
transformations. Described by the vector field with the same Lagrangian
(\ref{Lagrangian}), the dark matter is of the same physical nature in both
applications: to cosmology \cite{Meierovich1}, and to galaxy rotation curves.

As a matter of fact, I agree with the Sanders' statement that
\textquotedblleft..the correct theory may well be one in which MOND reflects
the influence of cosmology on local particle dynamics and arises only in a
cosmological setting\textquotedblright\cite{Sanders}. However, it is evident
that I don't share the Sanders' conclusion: \textquotedblleft\ It goes without
saying that this theory is not General Relativity, because in the context of
General Relativity local particle dynamics is immune to the influence of
cosmology\textquotedblright\cite{Sanders}. I have presented here the complete
derivation from the Einstein equations (\ref{0-0 Eq}%
-\ref{fi-fi and teta-teta Eqs}) to the galaxy rotation curve (\ref{V (r)}).

\ There are attempts of applying the scalar, vector, and tensor fields in
order \textquotedblleft to explain the flat rotation curves of galaxies and
cluster lensing without postulating exotic dark matter\textquotedblright%
\ \cite{Moffat}. In quantum physics each elementary particle is a quantum of
some field, and vice versa, each field corresponds to its own quantum particle
\cite{Rubakov}. From my point of view, the various fields are just convenient
mathematical instruments that we utilize for description of physical
phenomena, no matter how we name them.

According to the observations the period of oscillations $\frac{2\pi}{m}$ (see
Figures 6 and 7) is some $15$ kpc.\ If in quantum mechanics it is the de
Broglie wavelength $\lambdabar=\frac{\hslash}{mc}$, then the rest energy of a
quantum particle, corresponding to the vector field, should be $mc^{2}%
\sim\allowbreak2.\,\allowbreak5\times10^{-27}$ eV.

A few words about fine tuning. I have come across this situation for three
times. The first one has been the widely used in the fifties \textquotedblleft
Bennet pinch\textquotedblright\ \cite{Bennet} -- a fine tuned solution of
equilibrium of a high current channel where the magnetic attraction is
balanced by the gas pressure of electric charges. In reality it came out to be
a boundary between the expansion and compression when the balance is broken
\cite{Meierovich3}. For the second time it has been the conclusion of the
existence of the limiting mass of an ultra relativistic star by Chandrasekhar
\cite{Chandrasekhar}\ and Landau \cite{Landau}. The fine tuned solution of
equilibrium with ultra relativistic equation of state turned out to be the
boundary of the gravitational collapse \cite{Oppenheimer Snyder}. The third
time it has been the fine tuned singular cosmological solution by Friedman
\cite{Friedman}, Robertson \cite{Robertson}, and Walker \cite{Walker}. In the
thirties, dark matter had not been taken seriously. With account of dark
matter the FRW singular solution turned out to be a lower boundary of the
regular oscillating cosmological solutions \cite{Meierovich1}. In all the
cases the requirement of regularity rules out the problem of fine tuning.

From my point of view it is time to reconsider the equilibrium
\cite{Oppenheimer Volkov} and collapse \cite{Oppenheimer Snyder} of
supermassive bodies taking into account the dark matter.

\begin{acknowledgement}
I am thankful to K. A. Bronnikov, M. Yu. Kagan, and V. I. Marchenko for
discussions, and to E. R. Podolyak for the help with analysis of equations.
\end{acknowledgement}


\begin{thebibliography}{99}                                                                                               %


\bibitem {Oort 1924}J. H. Oort; Arias, B; Rojo, M; Massa, M (June 1924), "On a
Possible Relation between Globular Clusters and Stars of High Velocity", Proc
Nath Acad Sci U S A. 10 (6): 256--260.

\bibitem {Zwicky 1933}F. Zwicky, \textquotedblleft Die Rotverschiebung von
extragalaktischen Nebeln\textquotedblright, Helv. Phys. Acta, 6, 110--127,
(1933) . (Title in English: The red shift of extragalactic nebulae ).

\bibitem {V Rubin}V. Rubin, W. K. Ford, Jr (1970). "Rotation of the Andromeda
Nebula from a Spectroscopic Survey of Emission Regions". Astrophysical Journal
159: 379.V. Rubin, N. Thonnard, W. K. Ford, Jr, (1980). "Rotational Properties
of 21 Sc Galaxies with a Large Range of Luminosities and Radii from NGC 4605
(R=4kpc) to UGC 2885 (R=122kpc)". Astrophysical Journal 238: 471.

\bibitem {Milgrom}M. Milgrom, \textquotedblleft A modification of the
Newtonian dynamics: Implications for galaxy systems\textquotedblright,
Astrophys. J., 270, 371--383, 384--389, (1983); \textquotedblleft A
modification of the Newtonian dynamics as a possible alternative to the hidden
mass hypothesis\textquotedblright, Astrophys. J., 270, 365--370, (1983).

\bibitem {Bekenstein}Bekenstein, J.D. "Relativistic gravitation theory for the
modified Newtonian dynamics paradigm", Phys. Rev. D70, 083509 (2004). (http://arxiv.org/abs/astro-ph/0403694v6).

\bibitem {Sanders}Sanders, R.H. "A tensor-vector-scalar framework for modified
dynamics and cosmic dark matter". 2005, preprint (astro-ph/0502222).

\bibitem {Brownstein}J. R. Brownstein and J. W. Moffat (2006). "Galaxy
Rotation Curves Without Non-Baryonic Dark Matter". Astrophysical Journal 636
(2): 721. arXiv:astro-ph/0506370. Bibcode 2006ApJ...636..721B. doi:10.1086/498208.

\bibitem {Moffat}Moffat, J.W. \textquotedblleft Scalar-Tensor-Vector Gravity
Theory\textquotedblright. 2005, J. Cosmology Astropart. Phys., 05, 003
http://arxiv.org/abs/astro-ph/0412195 (astro-ph/0412195); preprint (gr-qc/0506021).

\bibitem {Fameay and McGaugh}Benoit Famaey and Stacy McGaugh. "Modified
Newtonian Dynamics (MOND): Observational Phenomenology and Relativistic
Extensions". http://lanl.arxiv.org/abs/1112.3960v2.

\bibitem {Meierovich1}B. E. Meierovich. \textquotedblleft Towards the theory
of the evolution of the Universe\textquotedblright. Phys. Rev. D 85, 123544 (2012).

\bibitem {Bogolubov Shirkov}N.N.Bogolubov and D.V.Shirkov. "Introduction to
the theory of quantized fields". \textquotedblright Nauka\textquotedblright,
Moscow 1976, page 35. (In Russian)

\bibitem {Meierovich2}B. E. Meierovich, "Vector order parameter in general
relativity: Covariant equations", Phys. Rev. D 82, 024004 (2010)

\bibitem {Landau-Lifshits}L. D. Landau and E. M. Lifshitz, Theoretical
physics, vol.2, Field theory, page 382, "Nauka", Moscow 1973.

\bibitem {Oppenheimer Volkov}J. R. \ Oppenheimer and G. M. Volkoff. "On
massive neutron stars". Phys. Rev. Volume 55, 374-381 (1939).

\bibitem {Oppenheimer Snyder}J. R. \ Oppenheimer and H. Snyder. "On continued
gravitational contraction". Phys. Rev. Volume 56, 455-459 (1939).

\bibitem {Rubakov}V. A. Rubakov. \textquotedblleft Large Hadron Collider's
discovery of a new particle with Higgs boson properties\textquotedblright%
\ Physics-Uspekhi, 55, 949--957 (2012).

\bibitem {Bennet}W. H. Bennet. Phys. Rev. V. 45, p. 890 (1934).

\bibitem {Meierovich3}B. E. Meierovich \textquotedblleft Toward the
realization of electromagnetic collapse\textquotedblright. Sov. Phys. Usp. 29,
506--529 (1986).

\bibitem {Chandrasekhar}S. Chandrasekhar. \textquotedblleft The maximum mass
of ideal white dwarfs\textquotedblright. Astrophys. J. 74, 81-82 (1931).

\bibitem {Landau}L. D. Landau. \textquotedblleft To the theory of
stars\textquotedblright. Phys. Zs. Sowjet. 1, 285 (1932).

\bibitem {Friedman}A. Friedman. \textquotedblleft Uber die Krummung des
Raumes\textquotedblright. Z. Phys. 10, 377-386 (1922). (Title in English:
\textquotedblleft About the curvature of space\textquotedblright).

\bibitem {Robertson}H. R. Robertson. \textquotedblleft Kinematics and world
structure\textquotedblright. Astrophys. J., 82, 248-301 (1935); 83, 187-201
and 257-271 (1936).

\bibitem {Walker}A. G. Walker. \textquotedblleft On Milne's theory of
world-structure\textquotedblright. Proc. London Math. Soc., 42, 90-127 (1936).
\end{thebibliography}
\end{document}